# Sonic interaction with a virtual orchestra of factory machinery

Laurent S. R. Simon*
LIMSI-CNRS
Orsay, France

Florian Nouviale+
INSA Rennes/IRISA-Inria,
Rennes, France

Ronan Gaugne+
Université Rennes 1/IRISA-Inria,
Rennes, France

Valérie Gouranton+
INSA Rennes/IRISA-Inria,
Rennes, France

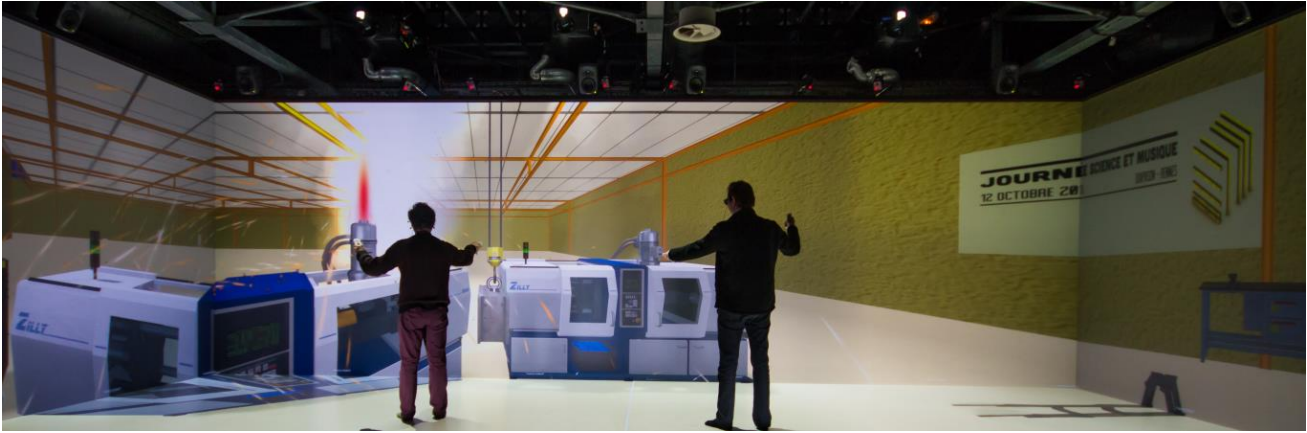

Figure 1: The virtual orchestra of factory machinery.

**ABSTRACT**

This paper presents an immersive application where users receive sound and visual feedbacks on their interactions with a virtual environment. In this application, the users play the part of conductors of an orchestra of factory machines since each of their actions on interaction devices triggers a pair of visual and audio responses. Audio stimuli were spatialized around the listener. The application was exhibited during the 2013 Science and Music day and designed to be used in a large immersive system with head tracking, shutter glasses and a 10.2 loudspeaker configuration.

**Keywords**: Sound spatialization, audio-visual interaction, virtual environments

**Index Terms**: H.5.2 [INFORMATION INTERFACES AND PRESENTATION (e.g., HCI)]: User Interfaces — *Auditory (non-speech) feedback*; H.5.5 [INFORMATION INTERFACES AND PRESENTATION (e.g., HCI)]: Sound and Music Computing— S*ignal analysis, synthesis, and processing*; H.5.5 [INFORMATION INTERFACES AND PRESENTATION (e.g., HCI)]: Multimedia Information Systems — *Artificial, augmented, and virtual realities*

## 1 INTRODUCTION

It has been proven that sound highly enhances the feeling of presence [1] in virtual reality environments [2]. In particular, in immersive structures such as CAVE [3] or assimilated facilities, audio rendering is carefully integrated in order to improve the feeling of being-there for the user, and several technologies have been proposed and studied to implement a rich sound environment [4][5]. Nevertheless, relatively little work explore the possibility for the user to dynamically create a sound universe in an immersive environment. For example, in [6] Berthault et al. propose an immersive application for 3D musical interaction. However, the interface presented to the user is very rich, with complex interactive interfaces and is mostly dedicated to trained musicians.

In this paper, we present an immersive environment for 3D sound interaction that was designed for a public exhibition around the theme "Science and Music", the *Journée Science et Musique* (JSM). The immersive application was designed to be used in Immersia (http://www.irisa.fr/immersia), a large CAVE-like infrastructure of the IRISA-Inria laboratory, located in Rennes, France. The aims of the application are (i) to immerse users in an interactive visual and sound environment, (ii) to be easy to access for the general public, and (iii) to illustrate in an entertaining way the use of sounds in VR.

The virtual environment used in this application places two users in a factory with two machines, close to them, as presented in Figure 1. By using Wiimote controllers, they can trigger and control punctual sounds, animations, sound and visual loops, as well as the global tempo of the application.

Section 2 presents the context in which this application was created. Then, section 3 describes the application and the hardware used. Finally, before concluding, Section 4 presents implementation details.

## 2 CONTEXT OF THE APPLICATION

### 2.1 The Immersia platform

The immersive platform of the IRISA-Inria computer science laboratory is a large virtual-reality facility dedicated to real-time, multimodal (vision, sound, haptic, BCI) and immersive interaction. It hosts experiments involving interactive and collaborative virtual-reality applications.

Images are rendered on four glass screens, a front one, two sides and a ground with an overall dimension of 9.6m wide, 3m deep and 3m high. Over 20 million pixels are displayed using a video projection system combining thirteen 3D HD Barco projectors. The tracking system is composed of 16 ART TRACK2 infrared cameras. Sound is spatially rendered using either a 10.2 sound

* laurent.simon@limsi.fr
+ firstname.lastname@irisa.fr

system with speakers or a 5.1 headset system, controlled by the user's position.

Immersia is a key node of the FP7 European Project Visionair [7] which goal is to create a European infrastructure that should be a unique, visible and attractive entry towards high-level visualization facilities for Virtual Reality, Scientific Visualization, Ultra High Definition, Augmented Reality and Virtual Services. These facilities, distributed across around twenty countries in Europe, are open and easily accessible to a wide set of research communities. Both physical access and virtual services are provided by the infrastructure. Full access to visualization-dedicated software is offered through call for projects, while physical access to high level platforms is partially accessible to other scientists, free of charge, based on the excellence of the project submitted. Immersia hosted three scientific projects from the beginning of Visionair, one in ergonomics, one in sport training, and one studying 3D sound perception in a large virtual environment.

### 2.2 The Science and Music day (JSM)

The JSM is an event of science popularization that occurs every year, during which a varied audience is invited to workshops, conferences and concerts on the theme of "Science and Music". The organizers attempt to keep a scope of activities as wide as possible, thus involving researchers, artists, engineers and technicians.

JSM has therefore become an event of discovery, meetings and discussions on the theme Science and Music.

More specifically, themes covered by the JSM include new interfaces for music interaction, room acoustics, music information retrieval, new medias, psychoacoustics, sound synthesis, 3D audio and virtual reality. In 2013, it included the Immersia workshop presented in this paper.

## 3 FACTORY CONCERT

### 3.1 Technical context and constraints

This project was conducted in a limited time. Consequently, it had to make use of existing spatialization tools instead of designing specific ones for Immersia. Ten Genelec 8040 and two Genelec subwoofers were available for use. They were linked to a Yamaha DME64N that performed the matrixing of the signals sent from the eight available outputs of the MOTU 896HD. Loudspeakers locations are shown on Figure 2. On this loudspeaker configuration, the two loudspeakers located at the center, front of the room, are connected to the same channel and placed 1.2m apart. Similarly, the two loudspeakers located at the center rear of the room, are connected to the same channel. It was not possible to replace them with a single loudspeaker, as the room is regularly used in dual mode, where each side of the room produces two independent images and sound workspaces. It was not possible either to use one different channel for each of the ten loudspeakers at the time of the project because of the limited number of free available inputs on the Yamaha DME64N.

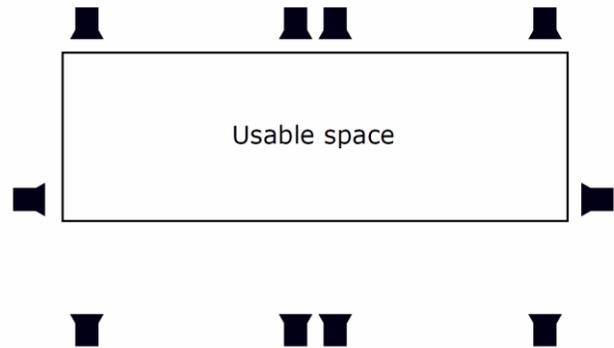

Figure 2: Location of the loudspeakers in the Immersia virtual reality room

The screens of the immersive room are made out of glass, and the images are retro-projected on them. The loudspeakers could therefore not be placed at ear height but had to be fixed above the 3 meters high walls. This could cause the sounds to be perceived as coming from above when the visual objects are located at ears height. In addition, it also means that the environment of the Immersia platform is highly reflective (RT60 = 1.30 second measured from the slope of the impulse response curve taken between 0.1 and 0.5 seconds), causing numerous strong early and late reflections. This should increase the Apparent Source Width (ASW) and decrease the locatedness of the audio sources, loudspeakers in this case [8][9]. It can be hypothesized that increasing the ASW of the loudspeakers would consequently increase the ASW of the sound images produced by these loudspeakers.

The surface of Immersia implies that the distance between adjacent loudspeakers is at least 1.94m, which can create spatial aliasing, as will be discussed in section 4.2.1.

The audio processing was performed in MAX/MSP. Aside of this software, all the other audio tools used were freely available tools.

As described in section 2.1, the position of the main user was tracked and the visual environment display was modified according to his point of view. This means that the system can be optimal for a single user only. Sound reproduction could therefore be optimized only for this single user as well. The system is therefore suboptimal for any additional user.

### 3.2 User interaction and sound feedback

The system is designed to be controlled by two users, both of them wearing shutter glasses, but only one of them tracked for visual and sound computations.

The tracked user is immersed in the virtual environment and can navigate inside this environment by walking inside the surface of Immersia. He or she is provided with two Nintendo Wiimote controllers. The buttons of the first Wiimote can start and stop two pairs of visual and audio loops on the main machine and trigger two pairs of animations and audio sounds. Using this Wiimote can also change the tempo. By shaking the second Wiimote, the user can trigger a pair of sound and visual effect.

The second user however is not immersed in the virtual environment. This is especially true when he is far from the tracked user and when this tracked user is close to one wall, where the display of the environment on the screens could differ and can be deformed a lot. However, as the scene is containing few objects and the relative position of the virtual scene displayed compared to the tracking origin was fixed, it is easy to interpret the display of the screens.

This second user is also provided with two Wiimote controllers. The buttons of first controller allow controlling the travelling crane by triggering sounds and animations making the crane move on a horizontal plane and following a rectangle path around one of the machines. By shaking this Wiimote, the user can trigger a pair of a maracas-like sound and a visual effect on the secondary machine. The buttons of the second Wiimote controller trigger other sounds and animations of the crane, which will move on the vertical axis. The crane can thus reach 3 different heights. By shaking the second Wiimote controller, the user can also trigger a pair of an udu-like sound and a visual effect at the same position as the first controller.

The buttons of all the Wiimotes triggered sound and visual effects played at the same speed while the shaking encouraged the users to follow the global tempo given by the other effects.

One of the uses of the Immersia platform is for archeological studies, where old rooms or buildings can be visualized and researchers can physically walk and navigate in the reconstructed place. One could therefore also imagine this immersive system being used for simulation of concert rooms. Provided it is possible to reproduce any auditory scene sufficiently faithfully in the Immersia platform, architects could walk in the concert room they designed and listen to how it sounds. They could also compare different solutions they might have for a concert room. Following this idea, we simulated two different rooms for the scene of this project: in one case, the simulated room had a Reverberation Time (RT) of 1.2 second; in the other, it was a church that had a RT of 7 seconds. The change of room was both visible and audible and was triggered by the user's position: when the main user was in the left half of the Immersia platform, the simulated room was the low RT environment, and when the user was in the right half of the Immersia platform, the simulated room was the church (see Figure 3). The positions of the scene object did not change, only the room did.

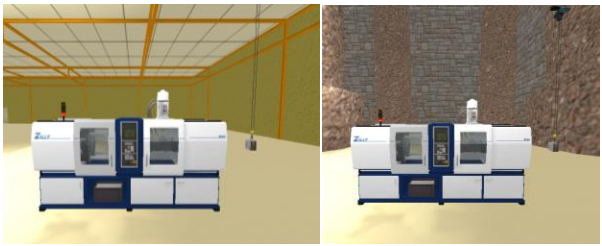

Figure 3: Factory / Church environment

The sounds used for the scene were recorded in a room that had an RT60 of 380ms. In order for the experiment to remain entertaining and amusing, it was chosen to record beatbox sounds. Two percussive sounds taken from Logic Pro 9's drums virtual instruments were added later on. The beatbox sounds included two loops recorded at 120 Beats Per Minute.

## 4 IMPLEMENTATION

The system, as shown on Figure 4 is based on a Max/MSP patch for the sound computation and a Unity3D based application with the MiddleVR plugin for user inputs, distribution and visual rendering on the screens. MiddleVR launches one application per rendering node, configuring each one according to a previously defined configuration file, and synchronizes all the instances. It also acts as a VRPN client on one node (called server) that broadcasts the inputs to the other nodes (called clients) so that every node handles the inputs at the same frame. The ART Tracking software and Nintendo Wiimote controllers are configured as input devices on the VRPN server accessed by the MiddleVR server application. This server node application is also in charge of sending data to Max/MSP through a UDP socket to update the virtual objects and user positions and trigger sound effects and tempo changes depending on the users' Wiimotes inputs.

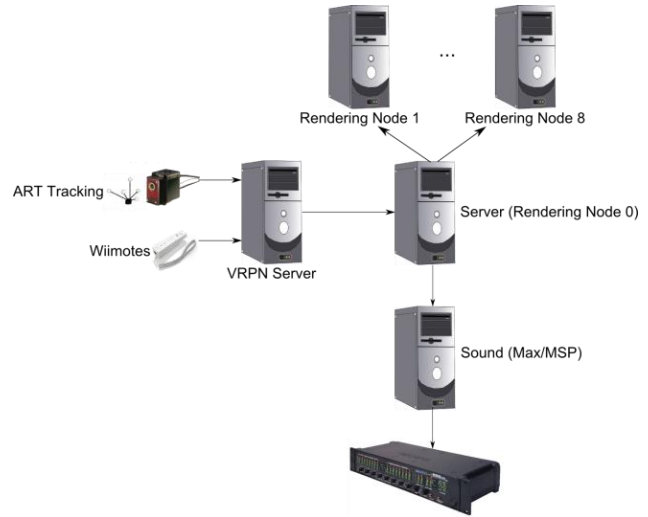

Figure 4: Hardware Architecture of the system

### 4.1 User interaction

To control the application, the two users are provided with two Nintendo Wiimotes each. This remote controller allows wireless interaction and presents buttons and sensors that part of the audience is already familiar with. Some of the buttons triggered events, resulting in sound and visual effects. The controls where defined as follows:

- The plus and minus buttons of the first Wiimote changes the tempo (i.e. the speed of the visual and sound effects, hence changing the pitch of the sounds)
- The directional pad of the first Wiimote can triggers punctual sound and visual effect. The A and up buttons enable or disable looping animations and sounds played at the global tempo
- The directional pad of the third Wiimote allows the second user to move the travelling crane (and therefore its sound source), triggering a sound and following a rectangle path on a horizontal plane
- The up and down buttons of the fourth Wiimote allows the user to move the travelling crane and its sound source up and down between three different heights.

We used the accelerometer of some of the Wiimotes to provide additional controls to the users. Thus, when the up axis value of the accelerometer reaches a threshold (i.e. when the user shakes the device from up to down), a sound and a visual effect are triggered. The visual effects consist of a short time emission of particles. The Wiimote accelerometers values correspond to the following rule: a value of one represents the standard gravity, minus one means the accelerometer is upside down. A value outside the range [-1;1] indicates that the user is shaking the controller in this accelerometer's axis.

The main user's shutter glasses are equipped with ART markers providing the application with the position and orientation of the user's head. This information is necessary to MiddleVR to compute the point of view to display on each screen but was also used as an input to the Max patch.

When the tempo changes are triggered by the plus and minus buttons of the first Wiimote the sounds, visual loops and animations times are divided by $2^{12}$, resulting in a pitch change of a semitone.

### 4.2 Spatialized sound

#### 4.2.1 Simulation of sound images' directions

In an immersive system, the notion of sweet spot is non-existent: subjects can move around freely. The sound reproduction system should therefore account for that. Since the position of the main user is tracked, in Immersia, binaural reproduction might seem like a good approach [12]. Binaural sound reproduction aims to simulate natural hearing by emulating the acoustic filters from a sound source to both of a listener's ears [8]. Sound is then reproduced over headphones and headtracking was shown to improve the quality of the rendering [11]. However, during the JSM, many users would be there to see the demonstration at the same time. Although only one of them can visualize the 3D content, we wanted all of them to be able to listen to the music that would be created by the main users. Time was also a constraint for the demo; equipping all users with headphones and headtrackers would therefore have been an issue. For these reasons, binaural would have been an ill-adapted sound reproduction technique for this project.

Another common approach to sound reproduction in immersive systems is to use sound field reproduction methods. These methods, namely Ambisonics and Wave Field Synthesis (WFS), can reproduce a sound field in a limited area [13][14] In an immersive system, it means that the users can walk around freely and still be able to localize any sounds of the auditory scene where they are supposed to. However, sound field reproduction techniques require loudspeakers to be numerous and placed close to each other. This is not possible in Immersia, as the surface to be covered by the loudspeaker system is large and the number of loudspeakers is small. According to [10], in WFS, the frequency $f_{al}$ above which spatial aliasing occurs is given by $f_{al} = c/(2d)$, where d is the distance between the acoustic centers of two adjacent loudspeakers. With a distance of at least 1.94m in Immersia, spatial aliasing would occur above 87 Hz. We considered using Ambisonics but did not have time to test it. Instead, the simple solution of Vector-Based Amplitude Panning was chosen [15].

The concept of VBAP is a simplification of Ambisonics: the level of each loudspeaker is controlled in order to make sure the level is constant at the listening position whatever the direction of the sound source is. Gains are applied to channels and the width of the source can be controlled by making more than two channels active at the same time. When a wide source is generated, the levels of the loudspeakers are higher the closer they are to the desired simulated position of the sound source. In this project, channel gains for the direct sound were given by VBAP with a source spreading width set to the minimum (0 out of 100):: the perceived source width was altered by the room effects. Despite its similarity with Ambisonics, it does not reproduce sound field over an extended area and therefore can only reproduce correct direction of sound sources at a given position, the sweet spot.

VBAP assumes all loudspeakers are located at the same distance from the listener. In order to estimate the gain necessary on each channel for a given sound source position, it needs the direction of each loudspeaker as well as the desired direction of the sound image. In Immersia, all the loudspeakers are not at the same distance from the listener, wherever the listener is. The position of the loudspeakers in the room is known, and so is the position of the listener. The position of the loudspeakers relatively to the listener can therefore be estimated continuously. The directions of the loudspeakers are corrected in VBAP. Figure 5 shows an example of how the loudspeaker angle may change relatively to the user.

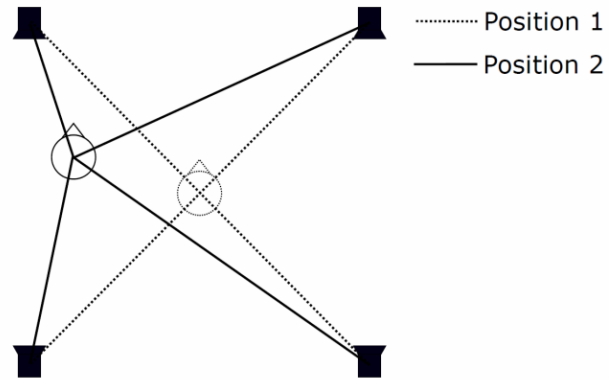

Figure 5: Modification of the loudspeaker angles when the listener moves. This figure shows two different possible listener positions in a simple loudspeaker configuration. It shows that the loudspeaker angles change in relation to the position of the listener.

For the room effect, we were limited to stereo reverberation processing: few multichannel reverberations exists for MAX/MSP, but they are developed either for 3-2 stereo systems (also known as 5.1 or ITU-R BSS.775-1 system) or convolution reverberation processors [16] Two-channel reverberation processors were therefore used for the room effects.

In room acoustics, early reflections and late reflections have very different perceptual effects: [17][9] show that early reflections improve the spatial impressions, increase the Apparent Source Width (ASW), and are perceptually merged with the direct sound. Late reflections, however, increase the perception of Listener EnVelopment (LEV) [18] and are directly perceived as a characteristic of the room. For geometrical reasons, early reflections tend to come more from directions close to the sound source than opposite the sound source, whereas late reflections are anisotropic. To simulate this property of room acoustics, direct sound, early reflections and late reflection were modeled and spatialized separately, as summarized on Figure 6.

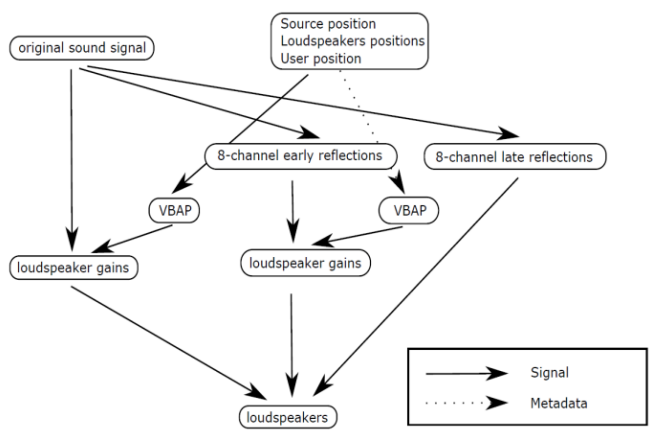

Figure 6: Routing of the audio signal in MAX/MSP

#### 4.2.2 Early reflections

In order to produce early reflections for eight independent channels, four stereo reverberation processes were used for each sound image, the direct and late reflections components being removed by adjusting the settings of the reverberation process. Using the same settings for each stereo process would have caused the four stereo room effects to be strongly correlated. According to [8], correlated sounds lead to a single perceived sound image, which is the opposite of what the early reflection processes are supposed to achieve here. The settings of each reverberation process were therefore altered. For the low RT room, the room size of each stereo reverberation process varied between 16 and 16.3 (no unit), while for the high RT room, the room size parameters varied between 143 and 143.3.

As explained above, early reflections are not anisotropic. This was simulated in Immersia by estimating channel gains in VBAP with a large source spreading width (50) and the same direction as the original sound source. For better spatial impression, early reflections should have come from each side of the direct sound, but not from the same direction as the direct sound.

#### 4.2.3 Late reflections

The late reflections are supposed to be anisotropic. The four reverberation processors were therefore uncorrelated in a similar way to the reverberation processors of the early reflections and then played to the same level on each of the channels, contrarily to early reflections. Since the source distance was not taken into account for the late reflections (in a diffuse field, the sound level is theoretically constant across space [19]), a single set of reverberation processors was used for all the sound images.

## 5 CONCLUSION

The Immersia JSM event fulfilled its three initial goals ((i) to immerse users in an interactive visual and sound environment, (ii) to be easy to access for the general public, and (iii) to illustrate in an entertaining way the use of sounds in VR.). The feeling of immersion, associated with intuitive interactions with the visual and sound environment provided the visitors an entertaining illustration of spatialized sounds in VR to which the public gave satisfied and encouraging feedbacks.

On the scientific side, this work addresses the implementation of real-time room acoustics tools in large immersive systems through loudspeakers.

One drawback of the application was the single valid point of view in the immersive structure, while two persons were interacting at the same time. There exist several solutions to bypass this problem, like using a combination of HMD, or putting the two users in two different CAVE-like systems. Some immersive structures offer several valid point of view at the same time by combining different stereoscopic technologies [20].

In future works, a more formal evaluation of the application could be performed on the perception of the sound sources localisation.

#### ACKNOWLEDGEMENTS

The authors wish to thank Astrée Rossetti for her work on the implementation, and Jules Espiau for his kind support. They also want to thank all the volunteers and visitors of the event.